# Thermodynamics of static: Observers in Gibbs paradox

Liangsuo Shu[1,2], Xiaokang Liu[1], Suyi Huang[1], Shiping Jin[1,2,3]

[1]School of energy and power engineering, Huazhong University of Science & Technology. Wuhan, China.
[2]Innovation Institute, Huazhong University of Science & Technology. Wuhan, China.
[3]China-Europe Institute for Clean and Renewable Energy, Huazhong University of Science & Technology. Wuhan, China.

**Abstract**

In this work, Gibbs paradox was discussed from the view of observer. The limitations of a real observer are analyzed quantitatively. The entropy of mixing was found to be determined by both the identification ability and the information already in hand of an observer.

**Introduction**

Gibbs paradox is a famous issue about the entropy of mixing of ideal gas: the value remains unchanged regardless of the degree of difference between two gases, then vanishes once they are identical. Since involving the philosophical basis of statistical mechanics, since presented by Josiah Willard Gibbs in 1875 [1], it has received extensive attention of many researchers [2] including Landé [3], von Neumann [4], van Kampen [5] and Jaynes [6] and has yet to end [7–9].

One way to solve this paradox is giving the degree of difference a quantitative description and then introducing it into the entropy of mixing [3,4,7]. In the scheme of von Neumann [4], it is the overlap integral of the quantum states of two gases; in a recent scheme, Allahverdyan and Nieuwenhuizen [7] related it to "the possibilities of controlling the involved quantum states". However, it would be pointless to discuss the degree of difference if a specific observer is left out. Because, for an ideal observer (likened to God by van Kampen [5]), any slight difference can be recognized; for a completely disabled observer, on the other hand, even a huge difference cannot be identified. In this work, we will discuss Gibbs paradox from the view of observer. An information method that has been suggested by Jaynes [6] and tried by Tseng and Caticha [8] was used. The entropy of mixing was found to be determined by both the identification ability and the information already in hand of an observer. The result given by Gibbs is a particular solution of the general solution presented in this work.

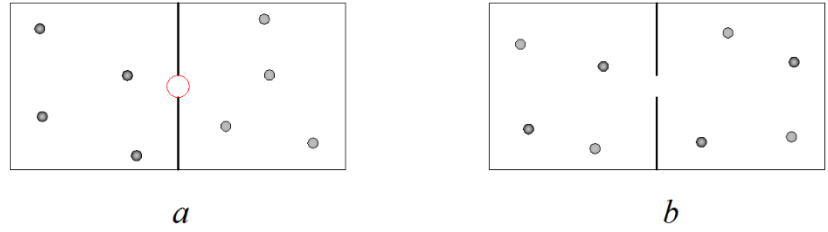

Fig.1 Schematic diagram of Gibbs paradox before mixing (*a*) and after mixing (*b*). Two kinds of balls with different grayscale represent different gases.

**Information and entropies from the views of different observers**

**Fig**.1 is a schematic diagram of Gibbs paradox. Two different gases are represented by balls with different grayscale: dark gray(*D*) and light gray(*L*), which are used to represent two gases below. Assuming there are 0.5 mole of *D* and *L* in the left and right chamber respectively. If an observer samples and analyzes the gas of two chambers before mixing, it will find both of them filled with a single kind of gas. The information ($I_0$) needed by the observer to describe the state is

$$I_0 = -0.5N_A \ln 1 - 0.5N_A \ln 1 = 0 \tag{1}$$

where $0.5N_A$ is the number of *D* and *L*, 1 is the probability of finding *D* or *L*. After the mixing, if the observer has the ability to distinguish *D* from *L* (**Fig**.2.*a*), the probabilities of finding *D* and *L* will both converge to 50% when the sample size is large enough, the information will change to $I_1$,

$$I_1 = -N_A \ln 0.5 = N_A \ln 2 \tag{2}$$

The information change is

$$\Delta I_{0 \to 1} = I_1 - I_0 = N_A \ln 2 \tag{3}$$

Entropy is a measure of system disorder. Boltzmann's entropy formula clarified the relationship between entropy of a macro-system(*S*) and the number of microstates (*W*),

$$S = k \ln W = kW \frac{1}{W}(-\ln \frac{1}{W}) \tag{4}$$

where *k* is the Boltzmann constant. According to the definition of information by Shannon, -ln(1/*W*) is the information needed to describe the uncertainty of a microstate with a probability of 1/*W*. After multiplying its probability 1/*W*, we will get the expectation of information to describe a particular microstate. As the number of microstates is *W*, therefore the total information needed to describe a state of the system($I_W$) will be,

$$I_W = W\frac{1}{W}(-\ln\frac{1}{W}) \qquad (5)$$

$I_W$ is also the information entropy defined by Shannon. Comparing **Eq**. (4) and **Eq**. (5), it is easy to get,

$$S = kI_W \qquad (6)$$

**Eq**. (6) clarifies the relationship between the entropy of a system and the information to describe it. Therefore, the entropy change ($\Delta S_{0\to1}$) between the two states from the view of an observer with the ability to distinguish $L$ from $D$ will be,

$$\Delta S_{0\to1} = k\Delta I_{0\to1} = R\ln 2 \qquad (7)$$

where $R$ is gas constant. If the observer cannot distinguish $L$ from $D$ at all (**Fig**.2.*b*), the probabilities of finding $D$ and $L$ will be either 0 or 1, the information will change to $I_2$,

$$I_2 = -N_A \ln 1 = 0 \qquad (8)$$

Therefore, for an observer without any ability to distinguish $L$ from $D$, the entropy change ($\Delta S_{0\to2}$) will be,

$$\Delta S_{0\to2} = k\Delta I_{0\to2} = 0 \qquad (9)$$

There is no heat effect in the mixing of two different ideal gases. Therefore, the calorimetric method is a completely disabled observer to distinguish ideal gases. In Gibbs paradox, there is a hypothesis hidden in the expression that the observer has a perfect identification ability for any slight difference. As a result, the entropy of mixing will be **Eq**. (7) for two different gases and change to **Eq**. (9) when two gases are undistinguishable. However, for a real observer, the identification ability of which is between an ideal observer and a completely disabled observer. Hence, the entropy change should be a value between the two limits described by **Eq**. (7) and **Eq**. (9). A real observer may make mistakes when distinguishing $L$ from $D$. Assuming the probability of its correct judgment is $P$, the probability distributions of $L$ and $D$ will be **Fig**.2.*c* or **Fig**.2.*d*: if the observer knows there are only two kinds of gas, it will take the unidentifiable $L$ as $D$ (**Fig**.2.*c*); if the observer do not have this information, it will take the unidentifiable $L$ as a third gas, N-$L$ (**Fig**.2.*d*). In the two cases, the information needed to describe the final state (after mixing) can be described by the following two equations respectively.

$$I_3 = -(1-0.5P)N_A \ln(1-0.5P) - 0.5PN_A \ln(0.5P) \qquad (10)$$

$$I_4 = -0.5N_A \ln 0.5 - 0.5PN_A \ln(0.5P) - (0.5 - 0.5P)N_A \ln(0.5 - 0.5P) \quad (11)$$

At the same time, the information needed to describe the initial state from the view of a real observer is also different from **Eq**. (1),

$$I_0' = -0.5N_A \ln 1 - 0.5N_A P \ln P - 0.5N_A(1-P)\ln(1-P) \quad (12)$$

The entropy changes from the view of a real observer will be,

$$\Delta S_{0' \to 3}/k = \Delta I_{0' \to 3} = 0.5PN_A \ln 2 + 0.5(1-P)N_A \ln(1-P) \\ -(1-0.5P)N_A \ln(1-0.5P) \quad (13)$$

$$\Delta S_{0' \to 4} = k\Delta I_{0' \to 4} = R\ln 2 \quad (14)$$

Comparing **Eq**. (7), **Eq**.(9), **Eq**.(13) and **Eq**.(14), it can be found that the entropy of mixing is determined by both the identification ability and the information already in hand of an observer.

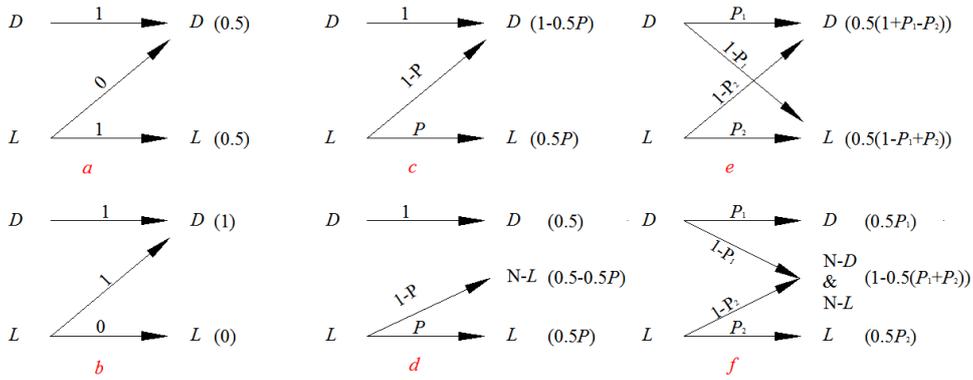

**Fig.2** Information and entropy from the views of different observers

In the above discussion, the accuracy of recognition of the real observer about $D$ is 100%. However, a real observer may also make mistakes in this recognition (**Fig**.2.$c$ or **Fig**.2.$d$). In these two cases, the information needed to describe the initial state will be,

$$I_0'' = -0.5N_A P_1 \ln P_1 - 0.5N_A(1-P_1)\ln(1-P_1) \\ -0.5N_A P_2 \ln P_2 - 0.5N_A(1-P_2)\ln(1-P_2) \quad (15)$$

the information needed to describe the final state will be,

$$I_5 = -(0.5P_1 + 0.5 - 0.5P_2)N_A \ln(0.5P_1 + 0.5 - 0.5P_2) \\ -(0.5P_2 + 0.5 - 0.5P_1)N_A \ln(0.5P_2 + 0.5 - 0.5P_1) \quad (16)$$

$$I_6 = -0.5P_1 N_A \ln 0.5P_1 - 0.5P_2 N_A \ln(0.5P_2) \\ -(1 - 0.5P_1 - 0.5P_2)N_A \ln(1 - 0.5P_1 - 0.5P_2) \quad (17)$$

The entropy of mixing will be

$$\Delta S_{0''\to 5} = k\Delta I_{0''\to 5} = k(I_5 - I_0'') \tag{18}$$

$$\Delta S_{0''\to 6} = k\Delta I_{0''\to 6} = k(I_6 - I_0'') \tag{19}$$

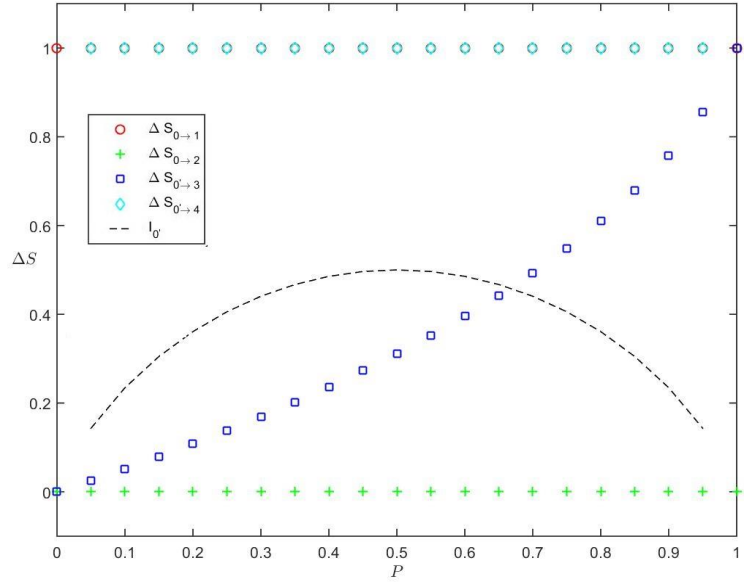

**Fig.**3 The entropy of mixing from the views of different observers which make mistakes in the recognition of $L$. The results are dimensionless by being divided by $R\ln 2$ for four values of the entropy of mixing or $\ln 2$ for $I_{0'}$. Four values correspond to $a$, $b$, $c$ and $d$ in **Fig.**2. $I_{0'}$ is the information needed to describe the initial state by these observers.

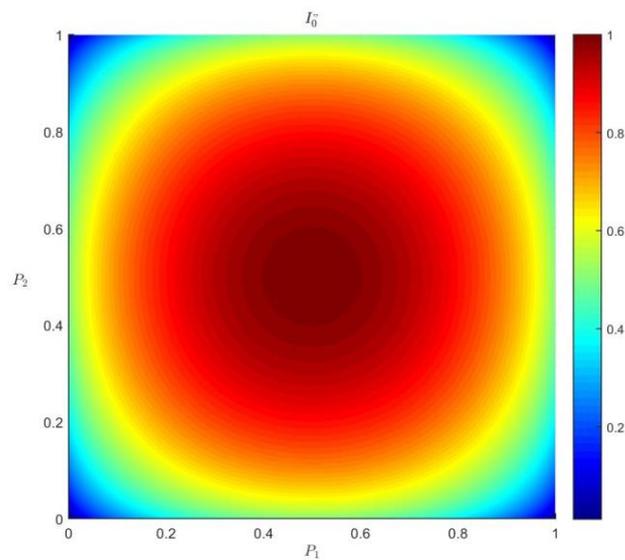

**Fig.**4 The information (/$\ln 2$) needed to describe the initial state by observer in $e, f$ in **Fig.**2

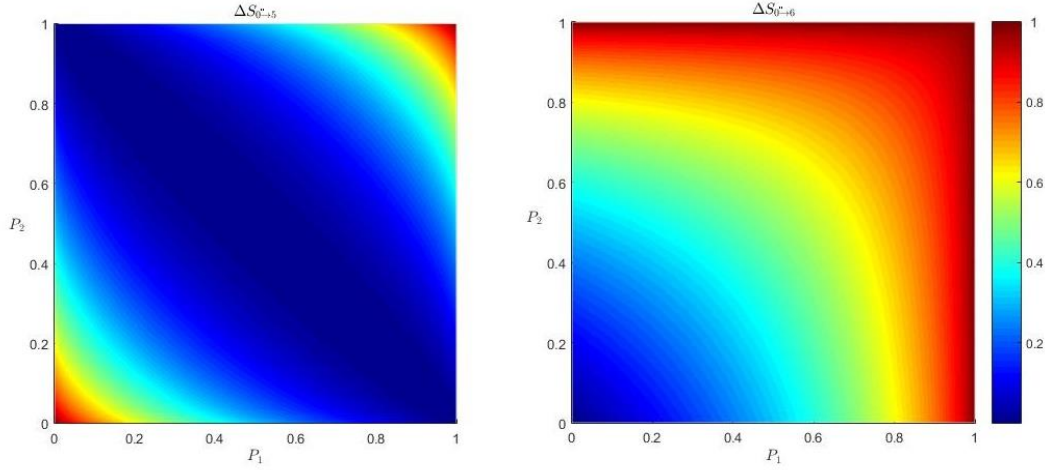

**Fig.**5 The entropy of mixing from the views of different observers which make mistakes in the recognition of both *L* and *D*. The results are dimensionless by being divided by *R*ln2. The values in two subgraphs (left and right) correspond to *e*, *f* in **Fig.**2

**Conclusion and Discussion**

In statistical mechanics of both classical and quantum, there is an implied ideal observer with perfect identification ability and complete information, which may not be, however, available to real observers. In this work, the defects in identification ability and incompleteness of information are taken into account. These two factors are found to affect the entropy of mixing.

The result given by Gibbs is a particular solution from the view of an ideal observer. Landé's [3] method implied the idea of real observer with imperfect identification ability. The "semipermeable diaphragms or other selective devices" in his argument can all be abstracted into the observer shown in **Fig**.2.*c*. However, he only realized the influence of the identification ability to the final state while ignoring its impact on the initial state and gave a result similar to **Eq**. (10).

In the discussion of different observers in **Fig.2**, they can only get the "correct" probability distribution of two gases when the sample size is large enough (the correctness has its meaning only for a particular observer since different observers may get different results). In practice, the sample size can also affect the results of a real observer.

In summary, the differences among real observers make them have different descriptions of a same process.

**Acknowledgements**

The authors thank Prof. Guoliang Xu for useful discussions and valuable suggestions. This work was supported by the National Natural Science Foundation of China (No. 51076057)